\documentclass[preprint,showpacs,preprintnumbers,amsmath,amssymb,nofootinbib]{revtex4}
\usepackage{booktabs}
\usepackage{mathrsfs}
\usepackage{epsfig}
\usepackage{graphicx}
\usepackage{dcolumn}
\usepackage{bm}
\usepackage{amsmath}
\usepackage{slashed}
\usepackage{multirow}
\usepackage{color}

\let\jnfont=\rm
\def\NPB#1,{{\jnfont Nucl.\ Phys.\ B }{\bf #1},}
\def\PLB#1,{{\jnfont Phys.\ Lett.\ B }{\bf #1},}
\def\EPJC#1,{{\jnfont Eur.\ Phys.\ Jour.\ C }{\bf #1},}
\def\PRD#1,{{\jnfont Phys.\ Rev.\ D }{\bf #1},}
\def\PRL#1,{{\jnfont Phys.\ Rev.\ Lett.\ }{\bf #1},}
\def\MPLA#1,{{\jnfont Mod.\ Phys.\ Lett.\ A }{\bf #1},}
\def\JPG#1,{{\jnfont J.\ Phys.\ G}{\bf #1},}
\def\CTP#1,{{\jnfont Commun.\ Theor.\ Phys.\ }{\bf #1},}
\def\ZPC#1,{{\jnfont Z.\ Phys.\ C }{\bf #1},}
\def\JHEP#1,{{\jnfont JHEP \ }{\bf #1},}
\def\Rv{\not{\hbox{\kern-1pt $R$}}}
\def\p{\not{\hbox{\kern-3pt $p$}}}

\newcommand{\bea}{\begin{eqnarray}}
\newcommand{\eea}{\end{eqnarray}}

\newcommand{\bcen}{\begin{center}}
\newcommand{\ecen}{\end{center}}

\newcommand{\ee}{e^+e^-}

\newcommand{\beq}{\begin{eqnarray}}
\newcommand{\eeq}{\end{eqnarray}}

\def\be{\begin{equation}}
\def\ee{\end{equation}}
\def\bea{\begin{array}}
\def\eea{\end{array}}
\def\beqa{\begin{eqnarray}}
\def\eeqa{\end{eqnarray}}
\def\beqas{\begin{eqnarray*}}
\def\eeqas{\end{eqnarray*}}

\def\bp{\begin{picture}}
\def\ep{\end{picture}}
\def\bc{\begin{center}}
\def\ec{\end{center}}
\def\bfig{\begin{figure}}
\def\efig{\end{figure}}

\def\bit{\begin{itemize}}
\def\eit{\end{itemize}}

\def\f{\frac}

\def\[{\left[}
\def\]{\right]}
\def\({\left(}
\def\){\right)}

\def\..{\left.}
\def\.{\right.}

\def\la{\leftarrow}

\def\la{\lambda}

\def\ep{\epsilon}

\def\pr{\prime}

\def\t1{\tilde{t_1}}

\begin{document}

\title{750 GeV diphoton resonance in a top and bottom seesaw model}
\author{ Archil Kobakhidze$^{1}$}
\author{ Fei Wang$^{2}$}
\author{ Lei Wu$^{1}$}
\author{ Jin Min Yang$^{3,4}$}
\author{ Mengchao Zhang$^{3}$}
\affiliation{$^1$ ARC Centre of Excellence for Particle Physics at the Terascale, School of Physics,
The University of Sydney, NSW 2006, Australia\\
$^2$ School of Physics, Zhengzhou University, Zhengzhou 453000, China\\
$^3$ Institute of Theoretical Physics, Academia Sinica, Beijing 100190, China\\
$^4$ Department of Physics, Tohoku University, Sendai 980-8578, Japan
   \vspace*{1.5cm} }

\begin{abstract}
The top and bottom seesaw model, which extends the top seesaw in order to accomodate a 125 GeV
Higgs boson, predicts vector-like top/bottom partners and these partners
can be bounded to form several neutral and charged singlet composite scalars by some new strong dynamics.
In this letter, we use such a singlet scalar to interpret the 750 GeV diphoton reseanance.
This singlet scalar is dominantly produced through the gluon fusion process induced by the
partners and its diphoton decay is induced by both the partners and the charged singlet scalars.
We show that this scenario can readily account for the observed 750 GeV diphoton signal under the
current LHC constraints. Further, this scenario predicts some other phenomenology, such as a strong
correlation between the decays to $\gamma\gamma$, $Z \gamma$ and $ZZ$, a three-photon singal from
the associate production of a singlet scalar and a photon, as well as some signals from the partner
cascade decays. These signals may jointly allow for a test of this framework in future 100 TeV hadron collider and ILC experiments.

\end{abstract}

\maketitle

\section{INTRODUCTION}
The observation of a 125 GeV Higgs boson at the LHC Run-1 \cite{higgs-atlas,higgs-cms} is a great triumph of the Standard Model (SM).
The current experimental measurements of its production cross sections and decay rates are consistent with the predictions of the SM
Higgs boson. However, without a symmetry protection, the SM Higgs mass is quadratically sensitive to the cutoff scale
via quantum corrections. This renders the SM rather unnatural and widely motivates new theories beyond the SM. Among many extensions
of the SM, the Higgs sector is usually enlarged or modified. So any evidence of non-SM Higgs bosons would indicate
the existence of new physics and can be used to elucidate the electroweak symmetry breaking (EWSB) mechanism.

Very recently, the ATLAS and CMS collaborations have reported their first results at 13 TeV LHC and found a resonance-like excess
in the diphoton invariant mass spectrum around 750 GeV \cite{atlas-diphoton,cms-diphoton}. The
significances of the signals are still only $3.6\sigma$ and $2.6\sigma$ in the respective experiments, but if confirmed with
more data, this would open the window of new physics at the TeV scale. Several explanations have been proposed for
such an excess \cite{ex-1,ex-6,ex-10}. When interpreting the excess in terms of the production rate of the resonance $X$,
based on the expected and observed exclusion limits,
the CMS and ATLAS experiments at 13 TeV LHC approximately give \cite{ex-6}
\begin{eqnarray}
\sigma^{750}_{\gamma\gamma}(\rm CMS) &=& \sigma(pp \to X) \times Br(X \to \gamma\gamma)=5.6^{+2.4}_{-2.4} fb, \\
\sigma^{750}_{\gamma\gamma}(\rm ATLAS) &=& \sigma(pp \to X) \times Br(X \to \gamma\gamma)=6.0^{+2.4}_{-2.0} fb.
\end{eqnarray}
Combined with the 8 TeV data \cite{aa1,aa2}, the diphoton excess contributing to the combined production rate is given by \cite{ex-6}
\begin{equation}
\sigma^{750}_{\gamma\gamma} = (4.4\pm 1.1) ~\rm{fb}~.\label{excess}
\end{equation}

Because of the Landau-Yang theorem \cite{ly}, the 750 GeV resonance $X$ can only be a spin-2 or spin-0 particle. However, a graviton-like spin-2 particle with an universal coupling is disfavored by the searches for the $jj$ \cite{jj}, $ZZ$ \cite{zz1,zz2} and $t\bar{t}$ \cite{tt1,tt2}  resonances. Besides, to enhance the diphoton rate, other SM decay modes of the heavy resonance have to be suppressed. So, the most economic way is to construct a theory with a spin-0 SM-singlet scalar $S$. Such a singlet naturally has no tree level couplings with the SM particles. While the large loop couplings $Sgg$ and $S\gamma\gamma$ can be achieved by introducing new vector-like fermions and/or new charged scalars, which can be found in some composite models and strong dynamics.

In this paper our aim is not to construct a full ultraviolet complete model, but instead work directly with an
effective framework inspired by the extension of the top seesaw with the bottom seesaw (namely top and bottom seesaw) \cite{cheng1,BLWY} to explain the 750 GeV diphoton
resonance without conflicting with other LHC data \footnote{The original top seesaw model can hardly explain
the 750 GeV diphoton excess since the mixing between top quark and top partner usually leads to a sizable
branching ratio of the resonance decay to $t\bar{t}$.}.
Because of the heavy mass ($m_t \sim 175$ GeV), top quark could potentially be associated with the EWSB.
The idea of top quark condensation was proposed to explain the EWSB, where a SM Higgs-like $t\bar{t}$
bound state (called the top-Higgs boson) with a mass $\sim 2m_t$ is predicted \cite{tcond}.
Obviously, the minimal top condensation model \cite{BHL} can hardly be consistent
with the recent measurements of the Higgs boson at the LHC. To accommodate 125 GeV Higgs boson, some extensions of top quark condensation with
seesaw mechanism \cite{topseesaw,HHT,fukano1,fukano2,fukano3,fukano4,fukano5,fukano6,cheng1,cheng2,BLWY} have been widely investigated.
Among them, top and bottom seesaw is a feasible way \cite{cheng1,BLWY}. Such models naturally predict the vector-like top and bottom partners,
which can be bounded to form several neutral and charged composite scalars by some new strong dynamics.
In our work we use such a neutral singlet scalar (composed of bottom partners) to interpret the
750 GeV resonance. This singlet scalar is dominantly produced through the gluon fusion process
induced by the partners and its diphoton decay is induced by both the partners and the charged
singlet scalars.
Under the current experimental constraints, we find that the 750 GeV diphoton excess can be
explained in this top and bottom seesaw scenario.

This paper is organized as follows. In Sec. \ref{sec2}, we describe the interactions relevant for the
750 GeV diphoton resonance and discuss the current experimental constraints. In Sec. \ref{sec3} we
present the numerical results. The conclusion is given in Sec. \ref{sec4}.

\section{The relevant interactions and constraints}\label{sec2}
We focus on the relevant interactions for the 750 GeV diphoton resonance within the framework of top and bottom
seesaw model \cite{BLWY}. Here we will concern only with the weak isospin singlet sector of the model and decouple it from the electroweak breaking sector that is assumed to correctly reproduce the observed Higss mass.
We start from the effective four-fermion
interactions (which are assumed to be generated by some strong dynamics at energy scale $\Lambda$) given by
 \beqa\label{fourfermion}
 {\cal L}_\Lambda \supseteq && [m_{0\chi}\bar{\chi}_L\chi_R +m_{0\omega}\bar{\omega}_L\omega_R +h.c.]+G_\chi(\bar{\chi}_L\chi_R)(\bar{\chi}_R\chi_L)+G_\omega(\bar{\omega}_L\omega_R)(\bar{\omega}_R\omega_L)\nonumber \\ &&+G_{\chi\omega}(\bar{\omega}_L\chi_R)(\bar{\chi}_R\omega_L)+G_{\omega\chi}(\bar{\chi}_L\omega_R)(\bar{\omega}_R\chi_L),
 \eeqa
where $\chi_{L,R}$ and $\omega_{L,R}$ are the vector-like top and bottom partners,
transforming as singlets under the electroweak $SU(2)_L$ gauge symmetry.
Their SM quantum numbers are given by
     \beqa\label{quark}
    \chi_L,\chi_R: (~3,~1,~2/3)~,~~~ \omega_L,\omega_R: (~3,~1,-1/3).
     \eeqa

At low energy scale $\mu$($<\Lambda$), the theory is described in terms of composite fields
corresponding to the bounded fermion pairs in Eq.(\ref{quark}). There are six composite
scalars relevant for our study, i.e., two neutral singlets $S_{N_i}$ and four charged singlets $S^{\pm}_{C_i}$ ($i=1,2$)\footnote{The masses of these composite singlets can be independent of each other since the global symmetry that protects the Higgs boson to be light, is imposed on the electroweak breaking sector \cite{cheng1,BLWY} and may be broken in the isospin singlet sector.}:
\beqa
S_{N_1} \sim \bar{\chi}_L \chi_R, \quad S_{N_2} \sim \bar{\omega}_L \omega_R,\quad\quad\quad\quad\quad\quad~~ \nonumber \\ S^+_{C_1} \sim \bar{\omega}_L \chi_R, \quad S^+_{C_2} \sim \bar{\omega}_R \chi_L, \quad S^-_{C_1} \sim \bar{\chi}_L \omega_R, \quad S^-_{C_2} \sim \bar{\chi}_R \omega_L.
\eeqa
Then, the effective Lagrangian describing the interactions between vector-like quarks and the scalars
as well as the self-interactions of the scalars can be written as
  \beqa
  \label{yukawa}
   {\cal L}_{\mu<\Lambda}\supseteq && y_{N_1} S_{N_1} \bar{\chi}_L\chi_R + y_{N_2} S_{N_2}\bar{\omega}_L \omega_R + y_{C_1} S^+_{C_1}\bar{\omega}_L \chi_R + y_{C_2} S^+_{C_2}\bar{\omega}_R \chi_L \nonumber \\ && +m_{0\chi}\bar{\chi}_L\chi_R +m_{0\omega}\bar{\omega}_L\omega_R +h.c.  +V(S_{N_i},S^{\pm}_{C_i}),
  \eeqa
where the bare mass terms $m_{0\chi}$ and $m_{0\omega}$ are allowed by the SM gauge symmetry. Using large $N_c$ fermion loop approximation \cite{BHL}, the Yukawa couplings $y_{N_i}$ at leading order can be estimated as
\begin{equation}\label{nc}
y_{N_i} \simeq \frac{4\pi}{\sqrt{N_c\ln(\Lambda^2/\mu^2)}}.
\end{equation}
These couplings tend to infinity at the compositeness scale $\Lambda$ due to the compositeness condition. For example, when $\Lambda=10$ TeV, $\mu=1$ TeV and $N_c=3$, Yukawa couplings $y_{N_i} \simeq 3.4$ are predicted. To obtain smaller $y_{N_i}$, the cut-off scale $\Lambda$ should be higher, (in this case, the theory will suffer from the fine tuning, but which is not the focus of this work.), e.g. $\Lambda=10^{12}$ TeV, $\mu=1$ TeV, then $y_{N_i}\simeq 1$. However, it is noted that for $\Lambda \gg \mu $, the fermion bubble approximation may not be accurate enough and the full one-loop RG equations are needed to be solved \cite{BHL}. The potentially large anomalous dimensions can drive large $y_{N_i}$ values at the compositeness scale down to substantially lower values at low energies \cite{cvetic}. Depending on the details of the full theory, one may in principle end up with hierarchically different Yukawa couplings $y_{N_i}$ as well. The exact numerical results can be worked out in the full theory, but this is beyond the scope of this paper. After spontaneous symmetry breaking, the vector-like quark masses are $m_\chi=y_{N_1}\langle S_{N_1}\rangle+m_{0\chi}$ and $m_\omega=y_{N_2}\langle S_{N_2}\rangle+m_{0\omega}$.
Since the couplings of vector-like quarks to the neutral composite scalars are not proportional to their masses,
we can separate the vector-like quark masses from the strength of the interaction $y_{N_i}$.
This feature can potentially enhance the effective couplings of $S_{N_i}gg$ and $S_{N_i}\gamma\gamma$.

Besides the vector-like quarks, these new charged scalars can contribute to the diphoton decay of $S_{N_i}$.
The relevant terms of the effective potential $V(S_{N_i},S^{\pm}_{C_i})$ in Eq.(\ref{yukawa}) are given by \footnote{In a full theory, the singlets $S_{N_i}$ may mix with the neutral components of the electroweak doublets. This mixing can be small because the vacuum expectation values $\langle S_{N_i}\rangle$ can be small and even vanishing. Similarly, $S_{N_1}H^\dagger H$ and $S_{N_2}H^\dagger H$ interactions that are respectively induced by $t-\chi$ and $b-\omega$ loops can be further suppressed by the large cut-off scale $\Lambda$ due to the twice transition of $t-\chi$ and $b-\omega$. In this case, the contribution of $S_{N_i} \to hh$ channel to the total decay width of $S_{N_i}$ can be negligibly small. We also require $m_{S^\pm_{C_i}}> m_{S_{N_i}}/2$ to kinematically forbid the decay channel $S_{N_i} \to S^\pm_{C_i}S^\mp_{C_i}$.}
  \beqa
  \label{potential}
   V(S_{N_i},S^{\pm}_{C_i}) \supseteq&&  \sum^2_{i=1}\frac{1}{2}m^2_{S_{N_i}} S^2_{N_i} +\sum^2_{i=1}\frac{1}{2}m^2_{S^\pm_{C_i}}S^+_{C_i}S^-_{C_i}+\sum^2_{i,j=1}\lambda_{C_{ij}}\mu^\prime S_{N_i} S^+_{C_j}S^-_{C_j},
  \eeqa
where $m_{S_{N_i}}$ and $m_{S^\pm_{C_i}}$ are the masses of the neutral and charged singlets, respectively. $\mu^\prime$ is the dimensional parameter and assumed to be 1 TeV. Similarly to Yukawa couplings $y_{N_i}$,
the trilinear coupling $\lambda_{C_{ij}}$ can be estimated at leading order through the fermion bubble approximation,
\begin{equation}\label{self}
\lambda_{C_{ij}} \simeq \frac{32\pi^2}{N_c\ln(\Lambda^2/\mu^2)},
\end{equation}
Again, when $\Lambda \gg \mu$, the trilinear coupling $\lambda_{C_{ij}}$ can be reduced. While the neutral scalars develop VEVs, the four-fermion interactions in Eq.(\ref{fourfermion}) resulting in charged scalars are assumed to be sub-critical, in order to avoid spontaneous breaking of $U(1)_{em}$.

In the top and bottom seesaw framework, the top/bottom quark masses are naturally reduced by the vector-like partners
$\chi/\omega$ via the seesaw in the top/bottom sector. The mass matrices of top and bottom sectors
are given by
\begin{gather}
  \begin{pmatrix} \bar{t}_L & \bar{\chi}_L \end{pmatrix}
  \begin{pmatrix} 0 & \mu_1 \\ m_{t\chi} & m_{\chi} \end{pmatrix}
  \begin{pmatrix} t_R \\ \chi_R \end{pmatrix} , \quad
  \begin{pmatrix} \bar{b}_L & \bar{\omega}_L \end{pmatrix}
  \begin{pmatrix} 0 & \mu_2 \\ m_{b\omega} & m_{\omega} \end{pmatrix}
  \begin{pmatrix} b_R \\ \omega_R \end{pmatrix},\label{phmass}
\end{gather}
where the entries $\mu_1$, $\mu_2$, $m_{t\chi}$ and $m_{b\omega}$ arise from some strong dynamics.
After diagonalizing the above mass matrices with seesaw condition $m_{\chi} \gg m_{t\chi}, \mu_1$
and $m_{\omega} \gg m_{b\omega}, \mu_2$, we obtain the physical top/bottom quark masses as
\beqa
&&m^{ph}_t \approx \frac{\mu_1 m_{t\chi}}{m_{\chi}}, \quad m_\chi^{ph}\approx m_\chi~~,\\
&&m_b^{ph}\approx \f{\mu_2 m_{b\omega} }{m_\omega}, \quad m_\omega^{ph}\approx m_\omega~~.\label{phymass}
\eeqa
The mixing angles between top/bottom quarks and their partners are given by
\begin{equation}
\sin^2\theta_{t} \approx \frac{m^{ph}_t}{m_{\chi}} ,\quad \sin^2\theta_{b} \approx \frac{m^{ph}_{b}}{m_{\omega}}.\label{mixing}
\end{equation}
From Eqs.(\ref{yukawa}) and (\ref{mixing}), we can see that the mixing angle $\sin\theta_t$ cannot be very small,
typically ${\cal O} (0.1)$ for a 10 TeV scale vector-like partner due to the large top mass. So, if $S_{N_1}$ serves as
the 750 GeV diphoton resonance, it will have a large branching ratio of $S_{N_1} \to t\bar{t}$, which has been tightly
constrained by the null result of the $t\bar{t}$ resonance search at the LHC Run-1. On the other hand, bottom quark has a
small mass and then the mixing angle $\sin\theta_b$ can be naturally small. Then, if $S_{N_2}$ is chosen as the
750 GeV diphoton resonance, it can easily satisfy the LHC dijet constraint. So, in our numerical study we
require $m_{S_{N_2}}=750$ GeV.

\section{Numerical Calculations and Results}\label{sec3}
\begin{figure}[ht]
\centering
\includegraphics[width=12cm]{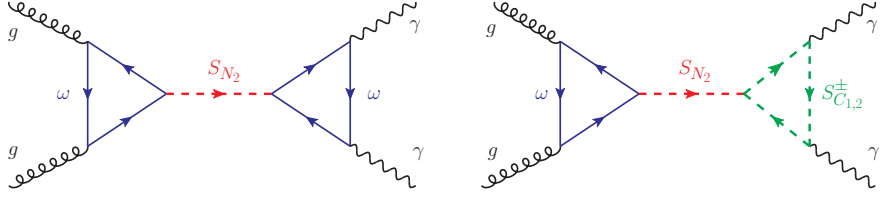}
\caption{Feynman diagrams for the process $gg \to S_{N_2}(750~{\rm GeV}) \to \gamma\gamma$.}
\label{feynman}
\end{figure}

In Fig.\ref{feynman}, we present the Feynman diagrams for the process $gg \to S_{N_2}(750~{\rm GeV}) \to \gamma\gamma$.
The gluon fusion production of $S_{N_2}$ is induced by the vector-like bottom partner $\omega$, while the diphoton decay
of $S_{N_2}$ is induced by both $\omega$ and the charged scalars $S^\pm_{C_{1,2}}$.
We calculate the production cross section of $gg \to h$ with $m_{h}=750$ GeV at the 13 TeV LHC by using the
package \textsf{HIGLU} \cite{higlu} with CTEQ6.6M PDFs \cite{cteq6}. The renormalization and factorization scales
are set as $\mu_R=\mu_F=m_S/2$. Then, the cross section of $gg \to S_{N_2}$ can be obtained as
$\sigma_{gg \to S_{N_2}}=(\Gamma_{S_{N_2}} \to gg / \Gamma_{h_{750}} \to gg) \cdot \sigma_{gg\to h_{750}} $. We also include a
$K$-factor $(1+67\alpha_s/4\pi)$ \cite{qcdcorrection} in the calculation of the decay width of $S \to gg$.

The main contributions\footnote{The bottom quark contributions is negligible small due to the tiny mixing
angle $\sin\theta_b$.} to the partial width of $S_{N_2} \to \gamma\gamma,gg$ are from the bottom partner $\omega$
and charged scalars $S^{\pm}_{C_{1,2}}$, which are given by
\begin{eqnarray}
  \Gamma_{\gamma\gamma} &\simeq& \frac{\alpha^2 m^3_{S_{N_2}}}{256\pi^3}\left|\frac{y_{N_2} \cos^2\theta_b}{m_\omega}N_c  Q^2_{\omega}A_{1/2}(\tau_\omega)+ \displaystyle{\sum_{i=1}^{2}} \frac{\lambda_{C_{2i}} \mu'}{2 m^2_{S^{\pm}_{C_i}}}  Q^2_{S^{\pm}_{C_i}}A_{0}(\tau_{S^{\pm}_{C_i}}) \right|^2, \\
  \Gamma_{gg} &\simeq& \frac{y_{N_b}^2 \cos^4\theta_b \alpha_s^2 m^3_{S_{N_2}}}{72\pi^3m^2_{\omega}}\left|\frac{3}{4}A_{1/2}(\tau_\omega)\right|^2 ,
\end{eqnarray}
where $N_c=3$ and $\tau_\omega=m_{S_{N_2}}^2/4m_{\omega}^2$, $\tau_{S^{\pm}_{C_i}}=m_{S_{N_2}}^2/4m_{S^{\pm}_{C_i}}^2$. The corresponding form factors of the fermion and scalar loops are
\begin{eqnarray}
  A_{1/2}(\tau)=2[\tau+(\tau-1)f(\tau)]\tau^{-2} \\
  A_{0}(\tau)=-[\tau-f(\tau)]\tau^{-2}
\end{eqnarray}
with
\begin{equation}\label{f}
  f(\tau)=
\begin{cases}
\arcsin^2\sqrt{\tau}& \text{$\tau\leq 1$}\\
-\frac{1}{4}\left[\log\frac{1+\sqrt{1-\tau^{-1}}}{1-\sqrt{1-\tau^{-1}}}-i\pi   \right]^2 & \text{$\tau>1$}
\end{cases}
\end{equation}

In our numerical calculations, the input parameters are $m_{S_{N_i}}$, $m_{S^\pm_{C_i}}$, $m_{\chi,\omega}$, $y_{N_i}$
and $\lambda_{C_{i1,i2}}$ ($i=1,2$). For simplicity, we assume the dimensionless parameters $\lambda_{C_{11}}=\lambda_{C_{12}}=\lambda_{C_1}$ and $\lambda_{C_{21}}=\lambda_{C_{22}}=\lambda_{C_2}$ and the charged singlet scalar masses $m_{S^\pm_{C_1}}=m_{S^\pm_{C_2}}=m_{S^\pm_{C}}$. The Yukawa couplings $y_{N_i}$ and trilinear couplings $\lambda_{C_{ij}}$ are expected to be large ($\sim O(1)$) but still perturbative ($\lesssim 4\pi$). Since the top partner sector will not contribute to the production rate of the 750 GeV diphoton resonance, we take $m_{S_{N_1}}=m_{\chi}=5$ TeV and assume the corresponding Yukawa coupling as $y_{S_{N_1}}=1$ and trilinear couplings $\lambda_{C_1}=2$ to avoid the constraints of the LHC search for $t\bar{t}$ high mass resonance \cite{tt1,tt2} and the electroweak precision observables on top partner sector \cite{fukano3}. It should be noted that the vector-like partner $\omega$ can provide the radiative corrections to the Peskin-Takeuchi parameter $T$ \cite{peskin}, and also the corrections to the $Zb\bar{b}$ vertex induced by the $b-\omega$ mixing in the bottom seesaw sector. According to Ref.~\cite{HHT}, the vector-like bottom partner $\omega$ should be heavier than 3 TeV to satisfy the requirement of these electroweak precision observables. Then, we scan the relevant parameters in the following ranges
\begin{eqnarray}
375 {~\rm GeV} \leq m_{S^\pm_{C}} \leq 2 {~\rm TeV},\quad  3 {~\rm TeV} \leq m_{\omega} \leq 10 {~\rm TeV}, \quad 1 \leq y_{N_2},\lambda_{C_2} \leq 4\pi.
\end{eqnarray}
Since there is no resonance observed in the searches for the $jj$ \cite{jj}, $ZZ$ \cite{zz1,zz2} and
$t\bar{t}$ \cite{tt1,tt2}, we impose the following constraints in the scan and require our samples to
explain the diphoton excess in the $2\sigma$ range of Eq.(\ref{excess}):
\begin{itemize}

\item [(1)] The CMS search for a dijet resonance~\cite{jj} at $\sqrt{s}=8$ TeV with ${\cal L} =18.8~fb^{-1}$ gives a $95\%$ C.L. upper limit on the production of the RS graviton decaying to $gg$,
\be
	\sigma(p p \to X)_{\rm{8\,TeV}} \times Br(X\to g g) < 1.8~\text{pb}~
	\label{ggbound}
\ee
\item [(2)] The ATLAS~\cite{zz1} and CMS~\cite{zz2} searches for a scalar resonance decaying to $VV(V=W,Z)$ at $\sqrt{s}=8$ TeV with the full data set, combining all relevant $Z$ and $W$ decay channels, give a $95\%$ CL upper limit on the production of the resonance decaying to $ZZ$,
\begin{align}
&\sigma(p p \to X)_{\rm{8\,TeV}} \times Br(X\to Z Z) < 22\,{\rm fb}\,_{\rm(ATLAS)} \,,~ 27\,{\rm fb}\,_{\rm(CMS)} \label{ZZbound}
\end{align}


\item [(3)] The ATLAS~\cite{aa1} and CMS~\cite{aa2} searches for a resonance decaying to $\gamma\gamma$ at $\sqrt{s}=8$ TeV give a $95\%$ CL upper limit on the production cross section:
\begin{align}
&\sigma(p p \to X)_{\rm{8\,TeV}} \times Br(X\to \gamma\gamma) < 2.2\,{\rm fb}\,_{\rm(ATLAS)} \,,~ 1.3\,{\rm fb}\,_{\rm(CMS)}\,.\label{aabound}
\end{align}

\item[(4)] The run-1 ATLAS resonance search in the $Z\gamma$ channel give a $95\%$ CL upper limit on the production cross section in the fiducial volume \cite{zabound}:
\begin{align}
&\sigma(p p \to X)_{\rm{8\,TeV}} \times Br(X\to Z\gamma) < 4 \,{\rm fb}\,_{\rm(ATLAS)}.\label{zabound}
\end{align}
\end{itemize}

\begin{figure}[ht]
\centering
\includegraphics[width=15cm]{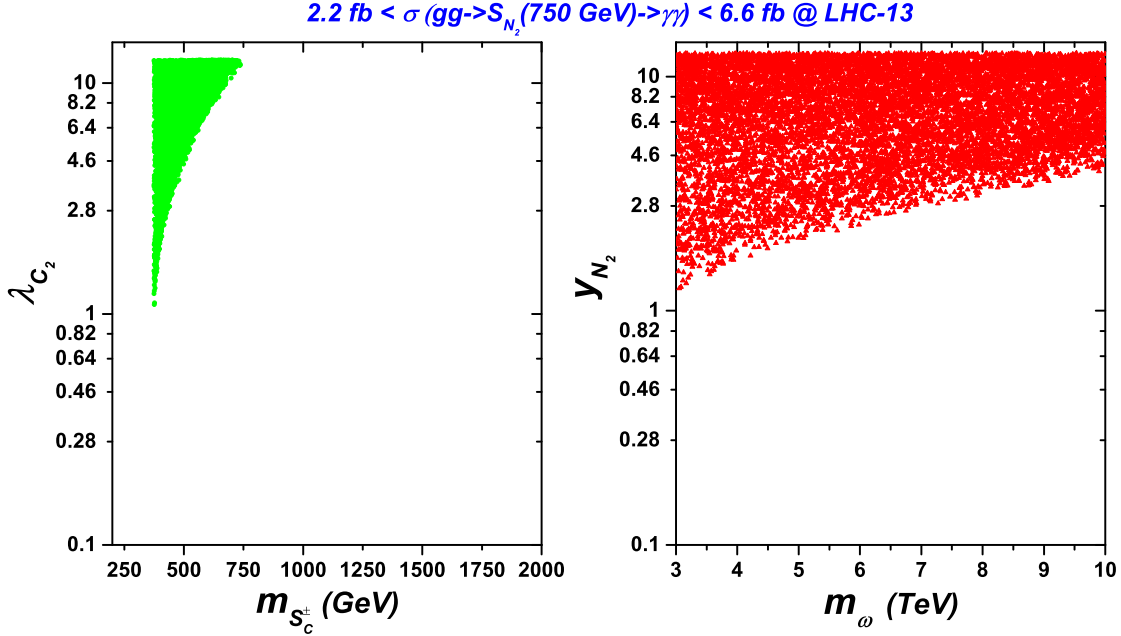}
\vspace{-0.5cm}
\caption{Scatter plots on the planes of $m_{S^\pm_{C}}$ versus $\lambda_{C_2}$, and $m_{\omega}$ versus $y_{N_2}$.
The samples satisfy the LHC constraints (1)-(5) and can explain the diphoton excess in the $2\sigma$ range
of Eq.(\ref{excess}). }
\label{result}
\end{figure}
In Fig.\ref{result}, we show the scatter plots on the planes of $m_{S^\pm_{C}}$ versus $\lambda_{C_2}$, and $m_{\omega}$
versus $y_{N_2}$. The samples are required to satisfy the LHC constraints (1)-(5) and to explain the diphoton excess in the
$2\sigma$ range of Eq.(\ref{excess}). We find that the most stringent constraints on the parameter space are from the
diphoton and dijet measurements at the 8 TeV LHC. The former imposes an upper limit on the cross section of
$gg \to S_{N_2} \to \gamma\gamma$ at the 13 TeV LHC, which is about 5.5 fb, while the latter tightly bounds the
total decay width $\Gamma_{tot}$ to be less than about 1 GeV \footnote{The compatibility of the signal hypothesis depends
on the width of the resonance. The CMS data prefers a narrow resonance, while the ATLAS local significance varies
from $3.6\sigma$ for a narrow width to $3.9\sigma$ for a broader resonance of $\Gamma_X/m_X \sim 6\%$. If the ATLAS
result persists with more data, the resonance should have a sizable width. In our model, the scalar resonance $S$
can also provide Majorana type mass for right-handed neutrino $N$ from the coupling $\la_N S \bar{N}^c N$. Then we
can have the TeV-scale seesaw mechanism in neutrino sector with tiny Yukawa coupling $y_\nu \bar{L}_L H N_R $.
If the Majorana mass for right handed neutrino is below 375 GeV, the scalar $S$ can also decay to neutrino pairs
and hence the total decay width can be enhanced. Another possibility is to introduce an additional singlet fermion
$N^\pr$. Then a natural TeV-scale inverse seesaw mechanism can be realized in the neutrino sector. The mass term
for $N^\pr$ can arise from the couplings $\la_{N^\pr} S N^\pr N^\pr$ and can be as light as 100 GeV \cite{neutrino}.
So, the scalar $S$ can also decay to $S^\pr$ pairs, leading to further enhancement of the decay width. All the additional
couplings involving $S$ can have a dynamical origin.  Assuming $N^\pr$ and $N$ share other non-critical gauge couplings
with $\chi$, the four-fermion interactions can generate the desired Yukawa couplings between $S$ and $N,N^\pr$
after the condensation of $\chi$. In all, the decay width of $S$ can be easily compatible with the ATLAS data in
our model.}. It is also worth mentioning that the bottom partner $\omega$ with the Yukawa coupling $y_{N_2}\gtrsim 1.24$
is mainly responsible for enhancing the cross section of the gluon fusion process $gg \to S_{N_2}$. Only $\omega$ propagating in the loop of $S_{N_2} \to \gamma\gamma$ can hardly explain the diphoton
excess since its electric charge is the same as bottom quark, $Q_{\omega}=-1/3$. On the other hand, the lighter charged scalars are, the smaller trilinear couplings are needed. So, from Fig.\ref{result} we can see
that the light strongly coupled charged scalars with $m_{S^\pm_{C}}\lesssim 750$ GeV and $\lambda_{C_2}\gtrsim 1.1$ are
needed to explain the diphoton excess. The heavier mass of $m_{S^\pm_{C}}$ is not viable due to the perturbative
requirement of the coupling $\lambda_{C_2}$.

In our scenario, besides the diphoton excess, other interesting signatures are predicted at the LHC and ILC.
For example, due to gauge symmetry, there is a strong correlation between the decay branching ratios of
$S_{N_2} \to \gamma\gamma$, $S_{N_2} \to Z\gamma$ and $S_{N_2} \to ZZ$, which is $1:2s_W^2/c_W^2:s_W^4/c_W^4$.
Moreover, the Drell-Yan process $q\bar{q}/e^+e^- \to S_{N_2}(\to \gamma\gamma)\gamma$ can produce three
distinctive hard photons in the final states and may be observed by the future LHC \cite{ramy} and ILC \cite{fujii}
experiments. Since the bottom partner $\omega$ is heavier than the 750 GeV resonance $S_{N_2}$, we can also have
the pair production of $\omega$, which leads to 2$b$-jets and four photons through the cascade decay process
$pp \to \omega\bar{\omega} \to S_{N_2}b S_{N_2}\bar{b} \to 2b+4\gamma$ at the future 100 TeV hadron collider. Given the current limited
significance of diphoton excess, with more data in LHC run-2, both ATLAS and CMS analyses will be able to
confirm this excess if it is indeed a signal of new physics beyond the SM. These signatures may be helpful
to further test our model at the LHC.

\section{conclusions}\label{sec4}
In this work, we interpreted the recent $\sim 750$ GeV diphoton excess at the 13 TeV LHC in a top and bottom seesaw model.
The neutral singlet composite scalar $S_{N_2}$ (composed of bottom partners) is chosen to play the role of the 750 GeV
resonance, which is dominantly produced through gluon fusion process $gg \to S_{N_2}$. Then, the diphoton decay rate
of $S_{N_2}$ can be greatly enhanced by the charged singlet composite scalars. We find that top and bottom seesaw model
can account for the observed 750 GeV signal without conflicting with other LHC constraints if the charged scalars
have the mass $m_{S^\pm_{C}}\lesssim 750$ GeV and the trilinear coupling $1.1 \lesssim \lambda_{C_2} \lesssim 4\pi$
and the bottom partner has the Yukawa coupling $1.24 \lesssim y_{N_2} \lesssim 4\pi$. Besides, this model predicts
other signatures, such as the strong correlation between
$S_{N_2} \to \gamma\gamma$, $S_{N_2} \to Z \gamma$ and $S_{N_2} \to ZZ$ decays, the three-photon signal
($q\bar{q}/e^+e^- \to S_{N_2}(\to \gamma\gamma)\gamma$) and the bottom partner cascade decay
($pp \to \omega\bar{\omega} \to S_{N_2}b S_{N_2}\bar{b} \to 2b+4\gamma$ ). If the diphoton excess is
further confirmed, these signatures may be helpful to test our model in future 100 TeV hadron collider and ILC experiments.

\acknowledgments
We thank Jing Shu for helpful discussions.
This work is partly supported by the Australian Research Council,
by the National Natural Science Foundation of China (NNSFC) under grants Nos. 11105124,11105125,11275057, 11305049,
11375001, 11405047, 11135003, 11275245,
by Specialised Research Fund for the Doctoral Program of Higher Education under Grant No. 20134104120002,
by the Open
Project Program of State Key Laboratory of Theoretical Physics, Institute of Theoretical Physics, Chinese Academy of
Sciences (No.Y5KF121CJ1), by the Innovation Talent project of Henan Province under grant number 15HASTIT017,
by the Joint Funds of the National Natural Science Foundation of China (U1404113),
by the Young-Talent Foundation of
Zhengzhou University, and by the CAS Center for Excellence in Particle Physics (CCEPP).


\begin{thebibliography}{99}

\bibitem{higgs-atlas}
  G.~Aad {\it et al.}  [ATLAS Collaboration],
  Phys.\ Lett.\ B {\bf 716}, 1 (2012).

\bibitem{higgs-cms}
  S.~Chatrchyan {\it et al.}  [CMS Collaboration],
  Phys.\ Lett.\ B {\bf 716}, 30 (2012).

\bibitem{atlas-diphoton}
 CMS Collaboration [CMS Collaboration],
  CMS-PAS-EXO-15-004.

\bibitem{cms-diphoton}
The ATLAS collaboration [ATLAS Collaboration],
ATLAS-CONF-2015-081.

\bibitem{ex-1}
  R.~Franceschini {\it et al.},
  arXiv:1512.04933 [hep-ph];
  S.~Di Chiara, L.~Marzola and M.~Raidal,
  arXiv:1512.04939 [hep-ph];
  K.~Harigaya and Y.~Nomura,
  arXiv:1512.04850 [hep-ph];
  A.~Angelescu, A.~Djouadi and G.~Moreau,
  arXiv:1512.04921 [hep-ph];
  Y.~Nakai, R.~Sato and K.~Tobioka,
  arXiv:1512.04924 [hep-ph];

\bibitem{ex-6}
  D.~Buttazzo, A.~Greljo and D.~Marzocca,
  arXiv:1512.04929 [hep-ph].

\bibitem{ex-10}
  T.~Higaki, K.~S.~Jeong, N.~Kitajima and F.~Takahashi,
  arXiv:1512.05295 [hep-ph];
   S.~D.~McDermott, P.~Meade and H.~Ramani,
  arXiv:1512.05326 [hep-ph];
  J.~Ellis, S.~A.~R.~Ellis, J.~Quevillon, V.~Sanz and T.~You,
  arXiv:1512.05327 [hep-ph];
  M.~Low, A.~Tesi and L.~T.~Wang,
  arXiv:1512.05328 [hep-ph];
  B.~Bellazzini, R.~Franceschini, F.~Sala and J.~Serra,
  arXiv:1512.05330 [hep-ph];
  R.~S.~Gupta, S.~Jger, Y.~Kats, G.~Perez and E.~Stamou,
  arXiv:1512.05332 [hep-ph];
  E.~Molinaro, F.~Sannino and N.~Vignaroli,
  arXiv:1512.05334 [hep-ph];
  D.~Curtin and C.~B.~Verhaaren,
  arXiv:1512.05753 [hep-ph];
  W.~Chao, R.~Huo and J.~H.~Yu,
  arXiv:1512.05738 [hep-ph];
  S.~Matsuzaki and K.~Yamawaki,
  arXiv:1512.05564 [hep-ph];
  Q.~H.~Cao, Y.~Liu, K.~P.~Xie, B.~Yan and D.~M.~Zhang,
  arXiv:1512.05542 [hep-ph];
  R.~Martinez, F.~Ochoa and C.~F.~Sierra,
  arXiv:1512.05617 [hep-ph];
  B.~Dutta, Y.~Gao, T.~Ghosh, I.~Gogoladze and T.~Li,
  arXiv:1512.05439 [hep-ph].

\bibitem{aa1}
CMS Collaboration [CMS Collaboration], CMS-PAS-HIG-14-006.


\bibitem{aa2}
G. Aad et al. [ATLAS Collaboration], Phys. Rev. D 92, 032004 (2015) [arXiv:1504.05511 [hep-ex]].


\bibitem{ly}
L. Landau, Dokl. Akad. Nauk Ser. Fiz. 60 (1948) 207; C. Yang, Phys. Rev. 77 (1950) 242.


\bibitem{jj}
  CMS Collaboration [CMS Collaboration],
  CMS-PAS-EXO-14-005.

\bibitem{zz1}
 G. Aad et al. [ATLAS Collaboration], arXiv:1507.05930 [hep-ex]; arXiv:1509.00389 [hep-ex].

\bibitem{zz2}
 V. Khachatryan et al. [CMS Collaboration], JHEP 1510, 144 (2015).

\bibitem{tt1}
G. Aad et al. [ATLAS Collaboration], JHEP 1508, 148 (2015).

\bibitem{tt2}
[CMS Collaboration], CMS-PAS-B2G-12-006.



\bibitem{cheng1} H.-C. Cheng, B. A. Dobrescu, J. Gu, JHEP08(2014)095

\bibitem{BLWY}   C.~Balazs, T.~Li, F.~Wang and J.~M.~Yang,
  JHEP {\bf 1301}, 186 (2013) [arXiv:1208.3767 [hep-ph]].



\bibitem{tcond} V. Miransky, M. Tanabashi, and K. Yamawaki, Phys. Lett. B221, 177 (1989);
   V. Miransky, M. Tanabashi, and K. Yamawaki, Mod. Phys. Lett. A4, 1043 (1989);
   W. Marciano, Phys. Rev. Lett. 62, 2793 (1989);
   W. J. Marciano, Phys. Rev. D41, 219 (1990);

\bibitem{BHL}
   W. A. Bardeen, C. T. Hill, and M. Lindner, Phys. Rev. D41, 1647 (1990).


\bibitem{topseesaw} B. A. Dobrescu and C. T. Hill, Phys. Rev. Lett. 81, 2634 (1998);
                 R. S. Chivukula, B. A. Dobrescu, H. Georgi and C. T. Hill, Phys. Rev. D 59, 075003 (1999);
                 B. A. Dobrescu, Phys. Rev. D 63, 015004 (2001).


\bibitem{HHT} H.-J. He, C. T. Hill and T. M. P. Tait, Phys. Rev. D 65, 055006 (2002).


\bibitem{fukano1}
  H.~S.~Fukano and K.~Tuominen,
  Phys.\ Rev.\ D {\bf 85}, 095025 (2012).
\bibitem{fukano2}
  H.~S.~Fukano and K.~Tuominen,
  arXiv:1210.6756 [hep-ph].
\bibitem{fukano3}
  H.~S.~Fukano and K.~Tuominen,
  JHEP {\bf 1309}, 021 (2013).
\bibitem{fukano4}
  H.~S.~Fukano, M.~Kurachi, S.~Matsuzaki and K.~Yamawaki,
  Phys.\ Rev.\ D {\bf 90}, no. 5, 055009 (2014).
\bibitem{fukano5}
  H.~S.~Fukano and S.~Matsuzaki,
  Phys.\ Rev.\ D {\bf 90}, no. 1, 015005 (2014).
\bibitem{fukano6}
  H.~S.~Fukano, M.~Kurachi and S.~Matsuzaki,
  Phys.\ Rev.\ D {\bf 91}, no. 11, 115005 (2015).

\bibitem{cheng2} H.-C. Cheng, J. Gu, JHEP10(2014)002.

\bibitem{cvetic}
  G.~Cvetic,
  Rev.\ Mod.\ Phys.\  {\bf 71}, 513 (1999).


\bibitem{higlu}
  M.~Spira,
  hep-ph/9510347.

\bibitem{cteq6}
  D.~Stump {\it et al.},
  JHEP {\bf 0310}, 046 (2003)
  [hep-ph/0303013].

\bibitem{qcdcorrection}
M. Spira, A. Djouadi, D. Graudenz and P.M. Zerwas, Nucl. Phys. B453 (1995) 17;
T. Inami, T. Kubota and Y. Okada, Z. Phys. C18 (1983) 69;
A. Djouadi, M. Spira and P.M. Zerwas, Phys. Lett. B264 (1991) 440;
M. Spira, A. Djouadi, D. Graudenz and P.M. Zerwas, Phys. Lett. B318 (1993) 347;
S. Dawson and R.P. Kauffman, Phys. Rev. D49 (1994) 2298.


\bibitem{peskin}
M. E. Peskin and T. Takeuchi, Phys. Rev. Lett. 65, 964 (1990);


\bibitem{zabound}
  G.~Aad {\it et al.} [ATLAS Collaboration],
  Phys.\ Lett.\ B {\bf 738}, 428 (2014).



\bibitem{neutrino} P. S. Bhupal Dev, Roberto Franceschini, R. N. Mohapatra,Phys. Rev. D86, 093010 (2012).

\bibitem{Elfgren:2000ch}
  E.~Elfgren,
  hep-ph/0105290.


\bibitem{ramy}
  R.~Brustein and D.~H.~Oaknin,
  Phys.\ Rev.\ D {\bf 62}, 015001 (2000);
  E.~Elfgren,
  hep-ph/0105290.

\bibitem{fujii}
  K.~Fujii, H.~Hano, H.~Itoh, N.~Okada and T.~Yoshioka,
  Phys.\ Rev.\ D {\bf 78}, 015008 (2008)
  doi:10.1103/PhysRevD.78.015008
  [arXiv:0802.3943 [hep-ex]].

\end{thebibliography}
\end{document}